\documentclass[sigconf]{acmart}

\AtBeginDocument{%
  }

\usepackage{listings}
\usepackage{booktabs}
\usepackage{tikz}
\usetikzlibrary{positioning}

\usepackage{subcaption}

\definecolor{codekeyword}{rgb}{0.0,0.0,0.55}
\definecolor{codecomment}{rgb}{0.4,0.4,0.4}
\definecolor{codestring}{rgb}{0.3,0.4,0.1}

\lstdefinelanguage{LispBM}{
  morekeywords={
    define,defun,lambda,let,progn,if,cond,match,
    recv,recv-to,spawn,spawn-trap,send,atomic,
    quote,and,or,loop,loopfor,loopwhile,looprange,
    loopforeach,loopwhile-thd,var,setq,set,
    trap,kill,sleep,yield,gc,self,closure,call-cc,
    image-save
  },
  sensitive=true,
  morecomment=[l]{;},
  morestring=[b]",
  alsoletter={-,!,?,*},
}

\lstdefinestyle{codebase}{
  basicstyle=\ttfamily\footnotesize,
  keywordstyle=\color{codekeyword}\bfseries,
  commentstyle=\color{codecomment}\itshape,
  stringstyle=\color{codestring},
  breaklines=true,
  breakatwhitespace=false,
  columns=fullflexible,
  keepspaces=true,
  showstringspaces=false,
  frame=single,
  framesep=2pt,
  xleftmargin=2pt,
  xrightmargin=2pt,
  tabsize=2,
  captionpos=b,
  numbers=left,
  numberstyle=\tiny\color{codecomment},
  numbersep=5pt,
}

\lstdefinestyle{cstyle}{
  style=codebase,
  language=C,
}
\lstdefinestyle{lbmstyle}{
  style=codebase,
  language=LispBM,
}

\lstnewenvironment{ccode}[1][]{\lstset{style=cstyle,#1}}{}
\lstnewenvironment{lbmcode}[1][]{\lstset{style=lbmstyle,#1}}{}
\lstnewenvironment{terminal}[1][]{\lstset{style=codebase,#1}}{}

\newcommand{\lbminline}[1]{\lstinline[style=lbmstyle]{#1}}
\newcommand{\cinline}[1]{\lstinline[style=cstyle]{#1}}

\setcopyright{acmlicensed}
\copyrightyear{2027}
\acmYear{2027}
\acmDOI{XXXXXXX.XXXXXXX}
\acmConference[IFL 2026]{38th Symposium on Implementation and
  Application of Functional Languages}{TBD 2026}{TBD}
\acmISBN{978-1-4503-XXXX-X/2026/TBD}

\title[Cases in Language Design]{A Linked List of Cases in Language Design}

\author{Bo Joel Svensson}
\authornote{Docent, unsalaried.}
\orcid{0000-0003-0363-1206}
\affiliation{%
  \institution{Chalmers University of Technology}
  \city{Gothenburg}
  \country{Sweden}
}
\affiliation{%
  \institution{VESC Labs AB}
  \city{Sexdrega}
  \country{Sweden}
}
\email{bo.joel.svensson@gmail.com}

\author{Benjamin Vedder}
\orcid{0000-0003-1713-3726}
\affiliation{%
  \institution{VESC Labs AB}
  \city{Sexdrega}
  \country{Sweden}
}
\affiliation{%
  \institution{Lind Art and Technology}
  \city{Stockholm}
  \country{Sweden}
}

\renewcommand{\shortauthors}{Svensson, et al.}

\begin{abstract}
  
  LispBM is a Lisp dialect for microcontrollers and embedded systems.
  This paper is about the changes made to LispBM resulting from it
  being integrated into the VESC family of firmware and thus gaining a
  userbase.

  As a surprise to Author 1, the new users of LispBM wrote programs
  larger than ever imagined. This used up all available heap space and
  we were forced to find a way to evaluate Lisp programs partitioned
  across RAM and flash memory. This story and a couple more,
  like the one about the Lisp-image system for fast boots, are the core
  contribution of this paper.

  When the first versions of the VESC firmware with a built-in Lisp
  runtime system were released, there was some grumbling here and there
  about Lisp and its syntax, as expected. To our delight, most seemed
  very excited about a scripting language in the firmware. To the
  engineer and builder users, the choice of syntax seemed to matter less
  than the potential it opened up for them.

\end{abstract}



\begin{document}

\maketitle

\section{Introduction}

LispBM is a dialect in the Lisp family of languages designed from the
beginning with microcontrollers as the primary target platform. Early
in the development of LispBM, design choices were driven by the 1st
author's interests and curiosity, first as a way to side-step the
time-consuming compile-flash-debug cycle of embedded systems
development. Later, while collaborating with Krook on
Scoria~\cite{Krook2022Scoria}, a language for real-time IoT
applications, and Sarkar on
Synchron~\cite{Sarkar2021SenseVM,Sarkar2022Synchron}, an API and
runtime for embedded systems, experience gained from building and
running LispBM fed directly into that work. The kinship with Synchron
runs deeper than a shared target platform: both systems ended up with
similar garbage collectors and heap designs. LispBM and Synchron chose different paths
on concurrency, however: Sarkar chose synchronous message passing
\`{a} la Concurrent ML. For LispBM we chose asynchronous message
passing with a pattern matching receive inspired by
Erlang~\cite{Armstrong2007History}.

In December of 2022 we started integrating LispBM into the VESC
(Vedder Electronic Speed Controller) motor control firmware. At the
time, author 1 expected this to be a short-lived experiment. LispBM,
however, has remained a part of the VESC firmware, and is by now
deployed in
hundreds, if not thousands, of electric vehicles ranging from hobbyist
machines to luxury range electric surfboards. With this change in
scope, LispBM was thrown into a situation where there are users,
requirements and drivers other than scientific curiosity.

This paper is about the choices, changes and problems that occur when
a particular language implementation project gains a userbase and a
whole new set of requirements compared to the premises it was built
around before. Here we present our experience in steering LispBM
to be the best language we can make it for its new role as a scripting
language embedded into the VESC family of firmware.

The next section, Section~\ref{sec:primer}, is a primer on what
LispBM programs and implementation look like. The sections that
follow the primer each tell a story of how the LispBM language
was changed, improved, or just pragmatically steered to align
with its new direction.

\section{LispBM: A Primer}\label{sec:primer}

The initial motivator behind LispBM was to get a
read-evaluate-print-loop (REPL) onto a microcontroller and in the
process learn about both Lisp implementation and microcontroller
limitations. LispBM does not attempt to be a Common Lisp~\cite{Steele1990CommonLisp} or a Scheme~\cite{SussmanSteele1975}
or compatible with any other pre-existing language in the Lisp family~\cite{mccarthy1960recursive},
rather we chose to be free to implement as we see fit in relation to
our needs and capabilities and the limitations of the target platform.

LispBM is a portable Lisp interpreter that you can slot into your
firmware on any platform that has a C compiler. Indeed, it is a
conscious choice not to include any platform-specific code into
LispBM. An example of this is that the LispBM
repository\footnote{\url{https://github.com/svenssonjoel/lispbm}}
holds a graphics library for microcontrollers equipped with screens, but no
drivers for any particular screen. The graphics library expects that a
platform that makes use of it will provide the appropriate callbacks.

\begin{figure*}
  \centering
  \includegraphics[width=\textwidth]{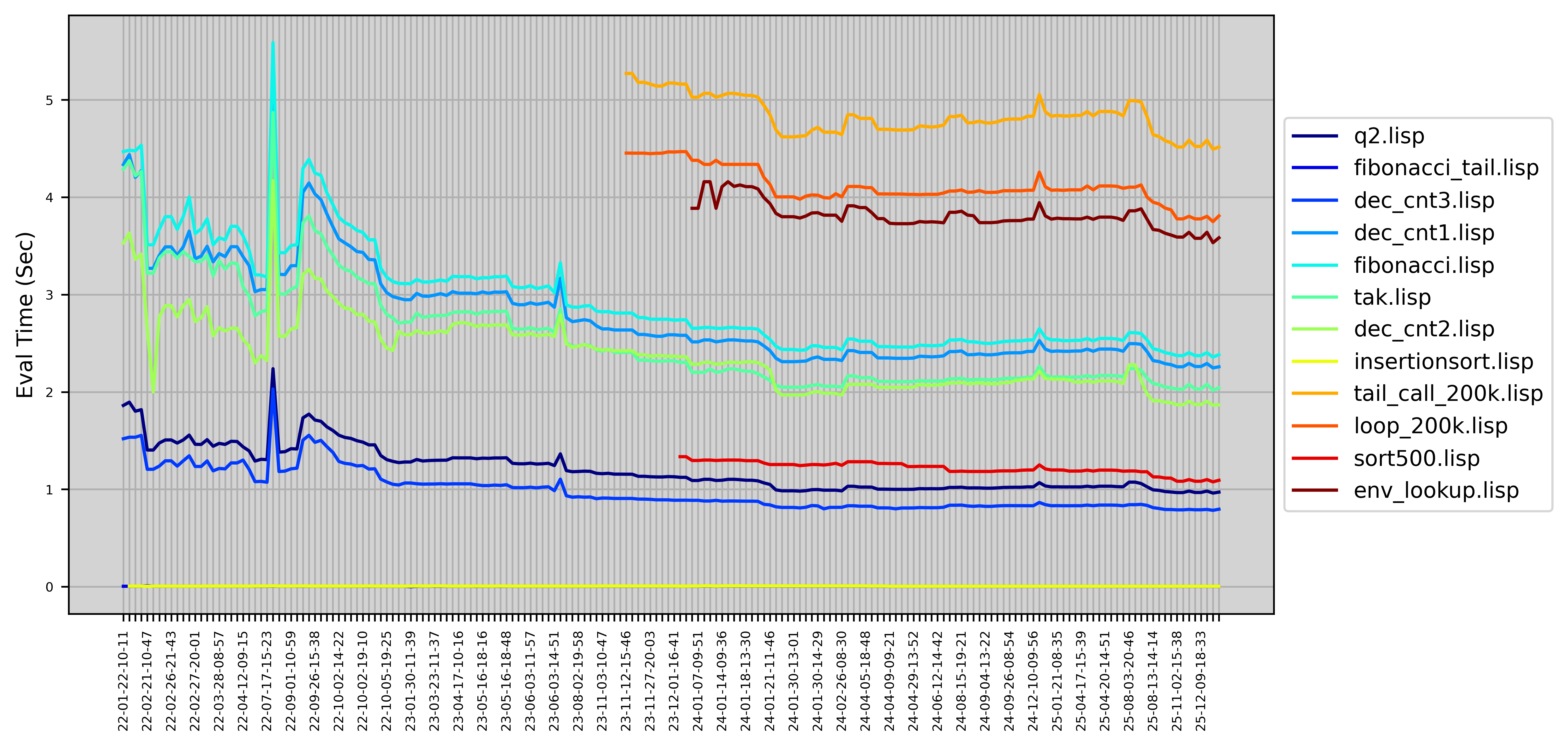}
  \caption{Evaluation time, in seconds, for a suite of benchmark
    programs run on an STM32F4-Discovery board. Performance has
    been tracked from January 2022 to August 2026.}
  \label{fig:benchresults}
\end{figure*}

Now, the goal for LispBM is to be a safe, correct, sandboxed runtime
system enabling scripting in mainly the VESC family of firmware.
Being a part of the VESC firmware places requirements on LispBM when
it comes to binary size, RAM usage and evaluation speed. We have
tracked the evaluator's performance on a set of benchmarks from
January 2022 up to August 2026. Figure~\ref{fig:benchresults} shows
how performance has evolved over that time. The
\lbminline{tail\_call\_200k} benchmark shows that a loop using
tail-calls can reach at most around 44kHz, while
\lbminline{loop\_200k} can reach a max of around 52kHz. Any
computation added to the bodies of these loops will significantly
decrease the obtainable max frequencies. LispBM is, however, not meant
to be the language in which to implement low-level motor control
loops. It is for scripting applications and services around the core
functionality provided by the firmware. Control loops implemented in
LispBM typically operate at a few hundred Hz, and are often limited to
that update frequency by sleeping.
% show a control program. Maybe the wheelie control.

Evaluation speed is currently
considered good enough for the kind of applications LispBM is used
for: dashboard graphics, remote control software, battery management
systems and ``wheelie''-mode for electric motorcycles, see Section~\ref{sec:wheelie}.

\subsection{An example program: Wheelie mode}\label{sec:wheelie}

\urldef{\wheeliesrcurl}\url{https://github.com/vedderb/vesc_pkg/blob/9d58644bcbf022d284099f163fa3972ab023b1a2/vl_bike_39p/code_stm.lbm#L519}

{\em Wheelie mode} or {\em Wheelie assist} is an application that
helps the rider of an electric motorcycle achieve a stable wheelie,
riding on only the rear wheel or wheels. For a
one-wheel vehicle, the concept of wheelie is undefined.

The code below shows how a wheelie mode implementation operates. This
code has been adapted from vl\_bike\_39p in the vesc\_pkg repository,
which is a collection of functionality for electric motorcycles~\footnote{\wheeliesrcurl}.
In the original source of this code, the wheelie mode calculation is executed
as part of a 150Hz control loop but in the example below it is a standalone
function that we {\em simulate} for a small number of inputs.

The wheelie mode makes use of LispBM's
\lstinline[style=lbmstyle]|{..}| syntax which is a short-hand for
\lbminline{(progn ..)}. The code also uses \lbminline{(var ...)} syntax
which is a way to create local bindings within a \lbminline{progn} block.

First some constants: 
\begin{lbmcode}
(def whl-start 20.0)
(def whl-end 43.0)
(def whl-kd 0.005)
\end{lbmcode}

The \lbminline{whl-start} constant is the angle at which wheelie-mode starts to assist the
driver in maintaining wheelie. This is the minimum wheelie angle. \lbminline{whl-end} is the
maximum angle we do not want to surpass while wheeling. The \lbminline{whl-kd} value is used
to dampen fast changes in pitch, if the front wheel is very rapidly rising, \lbminline{whl-kd} specifies
how much of a rate dependent damper we apply (a counter force).

\begin{lbmcode}
(defun simulate-wheelie (val-thr pitch-deg pitch-rate) {
   (var whl-gain (clamp (- 1.0 (/ (- pitch-deg whl-start)
                                  (- whl-end whl-start)))
                         -1.0 1.0))

   (var pitch-damper (clamp (* pitch-rate whl-kd) -1.0 1.0))

   (if (< pitch-deg whl-start)
       {
         (setq pitch-damper 0.0)
         (setq whl-gain 1.0)
       }
   )

   (var whl-thr (+ (* val-thr whl-gain) pitch-damper))

   whl-thr
})
\end{lbmcode}

The \lbminline{simulate-wheelie} function takes inputs
\lbminline{val-thr}, a throttle value from an external throttle position sensor;
\lbminline{pitch-deg}, the vehicle's current angle from an Inertial Measurement Unit (IMU);
\lbminline{pitch-rate}, how fast the angle is changing, as measured by a gyro sensor.
To stay faithful to how the real application is written, the \lbminline{pitch-rate}
value from the gyro is negative when the front wheel is moving upwards.

The \lbminline{simulate-wheelie} function makes use of a \lbminline{clamp} function:

\begin{lbmcode}
(defun clamp (v min max)
    (cond
        ((< v min) min)
        ((> v max) max)
        (t v)
))
\end{lbmcode}

\lbminline{clamp} is implemented using LispBM's \lbminline{cond} which is a multicase conditional
operator. \lbminline{cond} evaluates to the first expression from the top whose condition
evaluates to a truthy value (in LispBM any value other than \lbminline{nil}).

To evaluate the \lbminline{simulate-wheelie} over a range of pitch angles and a
constant throttle input, we write the following function:

\begin{lbmcode}
(defun run-wheelie-sweep () {
   (loopforeach pitch '(10.0 15.0 20.0 25.0 30.0 35.0 40.0 43.0 46.0) {
      (var cmd-thr (simulate-wheelie 1.0 pitch 0.0))
      (print (str-join (list "pitch: "
                             (to-str pitch)
                             " cmd-thr: "
                             (to-str cmd-thr))))
   })
})
\end{lbmcode}

\lbminline{run-wheelie-sweep} loops over a range of pitch angles and prints the
resulting throttle command value that would have been passed forward to the throttle
control loop in the real version of this code. Note that \lbminline{pitch-rate} is
a constant zero and thus no dampening effect is visible from this sweep.

The result of a \lbminline{run-wheelie-sweep} can be seen below:

\begin{terminal}
pitch: 10.000000f32 cmd-thr: 1.000000f32
pitch: 15.000000f32 cmd-thr: 1.000000f32
pitch: 20.000000f32 cmd-thr: 1.000000f32
pitch: 25.000000f32 cmd-thr: 0.782609f32
pitch: 30.000000f32 cmd-thr: 0.565217f32
pitch: 35.000000f32 cmd-thr: 0.347826f32
pitch: 40.000000f32 cmd-thr: 0.130435f32
pitch: 43.000000f32 cmd-thr: 0.000000f32
pitch: 46.000000f32 cmd-thr: -0.130435f32
\end{terminal}

In the output above we can see that the final throttle command gets
modified depending on our angle. In the beginning the angle is below
20 degrees and the throttle is passed through unaffected. Later, at 25
degrees, we can see that the throttle is being reduced by wheelie
mode, and the closer we get to \lbminline{whl-end} the more diminished
the output throttle command is.

To see that the \lbminline{whl-kd} factor has an impact on the output
throttle command, we run a few additional experiments:
\begin{terminal}
# (simulate-wheelie 1.0 40 0.6)
> 0.133435f32
# (simulate-wheelie 1.0 40 -0.6)
> 0.127435f32
# (simulate-wheelie 1.0 40 0)
> 0.130435f32
\end{terminal}

In the output above we can see that if the pitch angle is rising or
falling, the rate at which it changes is used to nudge the throttle
value to stabilize the change.

The code in this example is mostly arithmetic and arithmetic looks
about the same in any Lisp-like language. One point worth noting is
the use of a \lbminline{loopforeach} which is one of LispBM's many
loop macros. Because LispBM is targeting very resource constrained
platforms it is very easy to run out of stack using recursion unless
it is tail-recursion. For convenience we supply a number of built in
iteration patterns as well as macros that expand to good
tail-recursive forms.

\subsection{An example program: Code server}\label{sec:codesrv}

\urldef{\codesrvsrcurl}\url{https://github.com/vedderb/vesc_pkg/blob/9d58644bcbf022d284099f163fa3972ab023b1a2/lib_code_server/code_server.lisp}

Code server~\footnote{\codesrvsrcurl} is a VESC Lisp library that
can run a code execution service on a microcontroller that receives and
executes code sent by other microcontrollers over CAN (Controller Area Network).

This example contains code that is run on different microcontrollers
connected over CAN. The server is run on one device and provides a
very general way for anyone on the CAN bus to request information (and
affect the state of the server device) via remote function calls.

On the server side a worker that is handling the remote code execution
requests is running: 
\begin{lbmcode}
(defun code-server-worker (parent)
   (loopwhile t {
      (var rx (unflatten (canmsg-recv 0 -1)))
      (var id (first rx))

      (send parent (list 'can-id id))

      (if (>= id 0)
          (canmsg-send id 1 (flatten (eval (second rx))))
          (eval (second rx))
      )
}))
\end{lbmcode}

The worker receives a value (expression) using \lbminline{canmsg-recv}. The
value is serialized using the \lbminline{flatten} function on the sender side
and needs to be de-serialized here, using \lbminline{unflatten}. The worker expects that
the value is a pair of a can-id and payload. The can-id can be negative
for a fire-and-forget request.

The \lbminline{code-server-worker} is a monitored thread. There is a
parent thread that stands by to restart the worker if it crashes. On
line 6, the worker sends the can-id of the requester to the monitor
thread to allow it to send back a status in case the worker fails
during processing of its request.

Then, if the request wants a reply, the payload is evaluated using \lbminline{eval}.
The result of evaluation is flattened and sent back to the requester.
If the code sent to the worker crashes and brings down the worker thread,
the monitor thread is notified with an \lbminline{exit-error} message in its inbox.

Next we look at how to start the code server and its monitor:

\begin{lbmcode}
(defun start-code-server () {
   (spawn 150 (fn () {
      (var last-id 0)
      (var respawn true)

      (loopwhile t {
         (if respawn
             {
               (spawn-trap "CodeSrv" code-server-worker (self))
               (setq respawn false)
             }
         )

         (recv ((exit-error (? tid) (? v)) {
                (setq respawn true)
                (if (>= last-id 0)
                    (canmsg-send last-id 1 (flatten 'eerror))
                )
            })
            ((can-id (? id)) (setq last-id id))
         )
      })
  }))
})
\end{lbmcode}

The \lbminline{start-code-server} function spawns a function on line 2
(with a 150-element continuation stack, see Section~\ref{sec:eval}). \lbminline{fn} is shorthand
for \lbminline{lambda} and creates an anonymous function.  The spawned
thread loops forever and starts the worker thread on the first
iteration or restarts it if it has died in any following iteration. The
monitor then waits for messages from the worker. It receives a can-id
that it uses to feed back error messages to the requester if the worker dies.
The \lbminline{recv} form in LispBM is blocking and it uses pattern matching
on the received messages inspired by Erlang message passing.

On the client side we can make remote code execution requests using \lbminline{rcode-run}
and fire-and-forget requests using \lbminline{rcode-run-noret}:

\begin{lbmcode}
(def code-server-mutex (mutex-create))
  
(defun rcode-run (id tout code) {
   (mutex-lock code-server-mutex)
   (canmsg-send id 0 (flatten (list (can-local-id) code)))
   (var res (match (canmsg-recv 1 tout)
                    (timeout timeout)
                    ((? a) (unflatten a))
            ))
   (mutex-unlock code-server-mutex)
   res
})

(defun rcode-run-noret (id code) {
   (mutex-lock code-server-mutex)
   (var res (canmsg-send id 0 (flatten (list -1 code))))
   (mutex-unlock code-server-mutex)
   res
})
\end{lbmcode}

On the client side a mutex protects against having more than one
outstanding request in flight at any time from a specific
device. Note that multiple devices on the CAN bus can safely make
parallel requests.

\lbminline{rcode-run} takes a server-can-id, a timeout
(\lbminline{tout}) and code to evaluate on the server. The client's
can-id and the code are paired up and flattened and then sent over for
processing. The client then waits (canmsg-recv is a blocking
operation) for a reply and pattern matches on the response to detect
a timeout or a result value. \lbminline{rcode-run-noret} does the same but
immediately returns after sending its request.

Now that we have seen some examples of LispBM code, the next sections
go into how the runtime system that evaluates them works.

\subsection{Evaluator}\label{sec:eval}
The most central part of LispBM is its evaluator. The evaluator is
written in continuation passing style (CPS) inspired by the {\em
  lisperator} blog~\cite{Bazon2012CPSEvaluator}.  Compared to the
presentation in {\em lisperator}, which is using JavaScript, LispBM
being implemented in C uses what is in essence a defunctionalised CPS
evaluator~\cite{Svensson2022Defunctionalization}.

A Lisp program such as \lbminline{(+ 1 2)} is evaluated by pushing an
\cinline{APPLICATION\_START} continuation onto the continuation stack,
together with its arguments \lbminline{(1 2)} and the environment in
which to evaluate them. Then the head of the program, \lbminline{+},
is evaluated. When \lbminline{+} has evaluated and we can see that it
is an applicable function, the \cinline{APPLICATION\_START}
continuation is popped and processed. Processing an
\cinline{APPLICATION\_START} continuation, in turn, leads to a
sequence of \cinline{APPLICATION\_ARGS} continuations created and
retired.

The evaluator {\em step} function drives evaluation of Lisp terms
and it either applies a continuation on an evaluated form or
dispatches to the evaluator for the current form.

\begin{ccode}[caption={The evaluator step function}, label={lst:eval-step}]
static void evaluation_step(void){
  eval_context_t *ctx = ctx_running;

  if (ctx->app_cont) {
    lbm_value k = ctx->K.data[--ctx->K.sp];
    ctx->app_cont = false;

    // If app_cont is true, then top of stack must be a
    // valid continuation!
    // If top of stack is not a valid continuation CRASH!

    lbm_uint decoded_k = DEC_CONTINUATION(k);
    continuations[decoded_k](ctx);
    return;
  }

  if (lbm_is_symbol(ctx->curr_exp)) {
    eval_symbol(ctx);
    return;
  }
  if (lbm_is_cons(ctx->curr_exp)) {
    lbm_cons_t *cell = lbm_ref_cell(ctx->curr_exp);
    lbm_value h = cell->car;
    if (lbm_is_symbol(h) &&
        ((h & ENC_SPECIAL_FORMS_MASK) == ENC_SPECIAL_FORMS_BIT)) {
      lbm_uint eval_index = lbm_dec_sym(h) & SPECIAL_FORMS_INDEX_MASK;
      evaluators[eval_index](ctx);
      return;
    }

    // At this point head can be anything. It should evaluate
    // into a form that can be applied (closure, symbol, ...)
    // though.

    lbm_value *reserved = stack_reserve(ctx, 3);
    reserved[0] = ctx->curr_env; // INFER: stack_reserve aborts context if error.
    reserved[1] = cell->cdr;
    reserved[2] = START;
    ctx->curr_exp = h; // evaluate the function
    return;
  }

  eval_selfevaluating(ctx);
  return;
}
\end{ccode}

The \cinline{evaluation\_step} function in Listing~\ref{lst:eval-step} operates on a context,
\cinline{eval\_context\_t}, object called \cinline{ctx}. The context
represents a running LispBM program. We will refer to evaluation
contexts as {\em threads} throughout the rest of the paper, as that
is the more familiar term for a unit of concurrent execution,
reserving {\em context}/\cinline{ctx} for when we specifically mean
the C-level \cinline{eval\_context\_t} struct. A thread contains, most
importantly the current, \cinline{curr\_exp}, expression and the
current, \cinline{curr\_env}, environment. The flag,
\cinline{app\_cont} signals that the step before this one fully
evaluated a form and a continuation should be applied. The
continuation to apply is found at the top of the context's
continuation stack. If the \cinline{app\_cont} flag is false, the
\cinline{evaluation\_step} looks at the \cinline{curr\_exp} and
selects the appropriate path of evaluation depending on the type or
shape of that expression. The \cinline{curr\_exp} can here be a {\em
  symbol}, a list (which are applications), or something that
evaluates into itself (like numbers, strings etc).  A symbol is
evaluated by \cinline{eval\_symbol} which will perform environment
lookup and try to resolve the symbol binding.  If the
\cinline{curr\_exp} is a list, the head of that list is inspected and
a {\em special-form-path} or a general application-path is picked for
the expression. Special-forms are built-in syntax of the language such
as \lbminline{define}. Finally, if no other path applied, the
evaluator calls \cinline{eval\_selfevaluating} on the expression.

The \cinline{evaluation\_step} is called in a loop and each {\em
  evaluator} or continuation returns back rather than deeply
recurses. The evaluator avoids growing the C call stack with program
recursion depth. Instead, growth happens on the continuation stack,
whose depth tracks the program's syntactic nesting rather than its
recursion depth. The CPS style evaluator also gives space-efficient
tail-calls.

The LispBM evaluator supports Call-CC, macros and quasiquotations and
thus many things that are built directly into the evaluator could have
been implemented as Lisp libraries. Among these things you can find
concurrency, pattern matching, message-passing, and sorting. We view
these things as central to LispBM and have opted to implement them
inside of the evaluator for efficiency. This trades some binary size
for some execution efficiency, but it also means that Lisp programs
that make use of such features do not need to load them as Lisp code.

\subsection{Extensions}\label{sec:extensions}

All embedded software needs some way to interface with the real world.
LispBM leaves this to the integrator (for example VESC) to provide, in
the form of extensions, the functionality needed to interface with the
rest of the firmware and the hardware platform itself.

An extension is a C function with the following type:
\begin{ccode}
lbm_value my_extension(lbm_value *args, lbm_uint argn)
\end{ccode}

Extensions are registered with the runtime system and stored
in an extension table. A symbol used as the name of the function
on the Lisp side is generated at the time of registration.

The code below associates the Lisp name \lbminline{my-ext} with the
extension function \cinline{my\_extension}.
\begin{ccode}
lbm_add_extension("my-ext", my_extension);
\end{ccode}

Extensions arguments are Lisp values, that is values containing their
own type as part of their value, and these are passed to the extension
over an array. The extension code decodes Lisp values into C values,
operates on the C values and then encodes a Lisp value
result. Extensions are called from the same thread that runs the
evaluator so no other Lisp thread or runtime system service (such as
garbage collection) runs in parallel with the extension: The values passed to the
extension are in no danger of being garbage collected while the
extension is still accessing them.

\subsection{Memory}\label{sec:memory}

The embedded systems and microcontroller target platform have influenced the
design of the memory subsystem in LispBM. First, LispBM has a
traditional Lisp {\em heap}, a collection of car-cdr pairs, {\em
  cons-cells} that are allocated using the \lbminline{cons} operation,
\lbminline{(cons a b)} creates a pair of a and b, \lbminline{(a
  . b)}. In such a pair a and b can be any Lisp value.

The heap is managed by a mark and sweep garbage collector, chosen for
its low resource usage (Section~\ref{sec:gc}). This matters on most
of the target platforms, which have no caches: a memory read is a
memory read, and two reads always cost twice as much as one. A simple
mark and sweep can also use every single allotted byte before it needs
to run, something a copying garbage collector cannot do. As heaps used in practice
are quite small, garbage collection takes very little time. To
quantify this, we ran the micro-benchmark in
Listing~\ref{lst:gc-bench} on two representative targets. Table~\ref{tab:gc-bench}
shows the results.

\begin{lbmcode}[float, caption={Micro-benchmark used to measure garbage-collection time}, label={lst:gc-bench}]
(defun repeat (f times)
    (if (<= times 1)
        (f)
        { (f) (repeat f (- times 1)) }
))

(defun make-tree (n)
    (if (= n 0)
        (rand)
        (cons (make-tree (- n 1))
              (make-tree (- n 1)))))

(define a-tree (make-tree 8))
(define b-tree (make-tree 8))
(define c-tree (make-tree 8))
(define d-tree (make-tree 7))

(defun run-gc () {
        (def start (systime))
        (repeat gc 25)
        (def res (secs-since start))
        (print (list "GC-time (ms):" (* (/ res 25) 1000)))
})

(looprange i 0 100 (run-gc))
(print "Done!")
\end{lbmcode}

\begin{table}
  \centering
  \caption{Average garbage-collection time measured by the
    micro-benchmark in Listing~\ref{lst:gc-bench}, run on two VESC
    firmware targets (cf.\ Table~\ref{tab:memory-sizes} for heap
    sizing).}
  \label{tab:gc-bench}
  \footnotesize
  \begin{tabular}{lrrr}
    \toprule
    Platform & GC time (ms) & Heap usage & Heap size (cells) \\
    \midrule
    bldc (STM32F4)            & 2.0 & 69\% & 2753 \\
    VESC Express (ESP32-C3)   & 1.1 & 74\% & 2560 \\
    \bottomrule
  \end{tabular}
\end{table}

Strings, byte-arrays and boxed values (values larger than what can be
represented in a cell on the heap) are allocated on a memory area we
call {\em lbm\_memory}. The lifetime of objects allocated on
\cinline{lbm\_memory} by Lisp programs is managed by the same garbage
collector. The garbage collector is aware of allocations on
\cinline{lbm\_memory} via a sentinel that is created on the regular
cons-cell heap. A string, for example, is represented by a cons-cell
\lbminline{(ptr\_to\_string . SYM\_ARRAY\_TYPE)}. Here
\lbminline{SYM\_ARRAY\_TYPE} is a symbol that cannot be created
programmatically by a malicious user.

Heap and \cinline{lbm\_memory} sizes for a selection of target
platforms are shown in Table~\ref{tab:memory-sizes}.

The LispBM runtime system allocates most of its internal data
structures from \cinline{lbm\_memory}. The continuation stacks
associated with evaluation contexts are also allocated from
\cinline{lbm\_memory}. Heap size + \cinline{lbm\_memory} size is a
good estimate for the total RAM usage of LispBM. In fact, there
is one array, the \cinline{extension\_table}, which is not
currently allocated from \cinline{lbm\_memory}. From Table~\ref{tab:memory-sizes}
we can see that the bldc firmware allots just about 50KB of memory for LispBM.

\begin{table}
  \centering
  \caption{Heap and \texttt{lbm\_memory} sizes across VESC firmware
    variants, as configured in each firmware's default build. ESP32
    based variants (VESC Express) size memory per chip target at
    runtime and grow the heap when WiFi and/or BLE are disabled. The
    sizes below are the worst case with both WiFi and BLE enabled.}
  \label{tab:memory-sizes}
  \footnotesize
  \begin{tabular}{lrr}
    \toprule
    VESC FW variant & HEAP (cells) & \texttt{lbm\_memory} (bytes) \\
    \midrule
    bldc (STM32F4)                    & 2753 & 28672  \\
    VESC Express (ESP32-C3/C6)        & 2560 & 32768  \\
    VESC Express (ESP32-S3)           & 4608 & 49152  \\
    VESC Express (ESP32-S3/C6, PSRAM) & 8192 & 524288 \\
    \bottomrule
  \end{tabular}
\end{table}

\subsection{Concurrency}

Concurrency in LispBM follows directly from the continuation passing
style evaluator. Between any two calls to the step function in
Listing~\ref{lst:eval-step}, one can replace the continuation stack,
and other per program/thread state, such as \cinline{curr\_exp} and
\cinline{curr\_env} with those of another thread and in that way
effect a context switch.  In Listing~\ref{lst:eval-step}, line 2,
\cinline{eval\_context\_t *ctx = ctx\_running;} the context for which
to evaluate a step is chosen. The LispBM scheduler selects which
thread the \cinline{ctx\_running} points to.

Evaluation contexts are created from Lisp programs by calling
\lbminline{spawn}. Spawn takes a closure and arguments to evaluate
in a new context, see Listing~\ref{lst:spawn1}.

\begin{lbmcode}[caption={Spawn a thread}, label={lst:spawn1}]
(defun f (x) (print (+ x 1)))

(spawn f 100)  
\end{lbmcode}

When evaluating the program in Listing~\ref{lst:spawn1}, \lbminline{(spawn f 100)}
returns a thread-id and the value 101 is printed.

Evaluating a \lbminline{spawn} expression allocates a new context
on the \cinline{lbm\_memory} and adds this context to a queue of
runnable contexts. 

Each thread has a mailbox and messages can be sent to a thread
using its thread-id as we have seen in Section~\ref{sec:codesrv}.
Mailboxes do not grow dynamically and apply a drop-the-oldest policy
to deal with a full mailbox.

Section~\ref{sec:consched} goes into more depth on the concurrency
and scheduling systems used in LispBM.

%%%%%%%%%%%%%%%%%%%%%%%%%%%%%%%%%%%%%%%%%%%%%%%%%%%%%%%%%%%%%%
%% Performance picture

\section{Concurrency and Scheduling}\label{sec:consched}

LispBM was a concurrent language from early on in its development. To
begin with, the concurrency was cooperative. Cooperative concurrency
puts scheduling in the programmer's hands; each thread runs
uninterrupted until it explicitly yields using a \lbminline{(yield us)}
command.

Cooperative concurrency was very appealing from a runtime system
implementation perspective when starting out. As each thread is
responsible for putting itself to sleep, the runtime system's role is
to just wake up a suitable next-run thread. In LispBM this was handled
using two queues, a {\em runnable}/{\em sleeping} queue and a {\em
    blocked} queue. Threads that wait for a message on a \lbminline{recv} place
  themselves on the blocked queue and threads that yield go to the
  runnable/sleeping queue. The runtime system found the next thread to
  run searching through the runnable queue for a thread that has slept
  its requested duration using timestamp comparisons.

There are numerous arguments for cooperative concurrency: no timer
interrupts are needed to drive the context switch rate; as all threads
atomically evaluate until yield, there is no need for locks/mutexes; there
is a predictability in interleavings as they can only happen at fixed positions;
and, finally, the simple implementation already discussed.

Now, LispBM is used by people who are not necessarily computer scientists
or for that matter ``programmers'' (at least not first). The users of
LispBM are ``builders'' and engineers. Cooperative concurrency places
all the complexity of concurrent programming directly on the user. No
concurrent program can be written without understanding
cooperative concurrency. A round-robin model of concurrency can be
approached with less background and over time one can get familiar with
the more intricate details such as mutexes, which most LispBM programmers
may actually never need. A round-robin scheduling model is the pragmatic
choice!

In the fall of 2022, LispBM switched to round-robin scheduling. Now,
when the scheduler selected a thread to run, it was given a number of
evaluator steps in quota. When the quota was consumed, the thread was
put back at the end of the runnable queue. Just like the cooperative
scheduler, the round-robin scheduler was based on two queues, but here
a runnable queue was used for all threads that can run right now,
stored in order of when they retired due to quota. The other queue
holds blocked and sleeping threads; when a thread unblocks or finishes
sleeping, it moves to the end of the runnable queue.

The main problem with step-based quotas in LispBM was that the steps
take a different amount of time to execute. One example of this is that
extensions (see Section~\ref{sec:extensions}) are arbitrary C functions
run by the same thread as the evaluator, that is, in one step. This
issue became increasingly disruptive as we began building graphical
user interfaces based on the VESC Express platform. The extensions
that shuffle lots of pixels onto a screen take a lot of time, and a
typical GUI likely has a thread that performs many drawing calls one
after another; this is the worst-case scenario for a step-based
scheduler, which would happily run a full quota of steps (e.g., 50)
consisting mostly of expensive drawing operations.

At the beginning of 2025, users of the graphics library were doing complicated
enough GUIs that the step-based quota was too much of a problem, and
we introduced the time-based scheduler that has remained in use since.
Note that the time-based scheduler cannot pre-empt a thread mid-step. So,
extension implementors should still keep extensions as short as possible.

\section{Garbage Collection}\label{sec:gc}

LispBM uses a mark and sweep garbage collector informed by Jones and Lins, ``Garbage
Collection''~\cite{jones1996garbage}. This type of garbage
collector was chosen for its simplicity and its low requirements in terms
of meta-data or additional storage. Meta-data in the mark and sweep case
is limited to (in the basic form) just a mark bit per heap cell.

A mark and sweep garbage collector is named after its two phases of
operation. First, values that are alive are marked. This process,
called the mark phase, recursively finds all values reachable from a
set of starting points, for example the environment. After all alive
cells have been marked, the sweep phase iterates over all cells that compose
the heap. At this time all cells with a mark have their mark cleared, and all
cells that have no mark are added to the free-list.

Up until recently, the mark algorithm in use in VESC firmware was
entirely based on an explicit stack of a fixed size. Take for example
the program below:

\begin{lbmcode}
(define my-list (list 'a 'b 'c))
\end{lbmcode}

\begin{figure}
  \centering
  \includegraphics[width=0.6\linewidth]{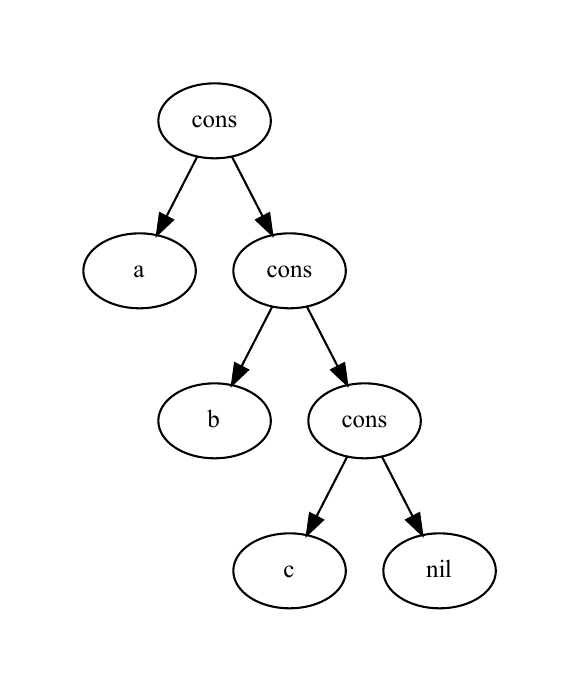}
  \caption{The heap structure built by \lbminline{(list 'a 'b 'c)}.}
  \label{fig:list1}
\end{figure}

After the program evaluates, the environment has a binding of the name
\lbminline{my-list} to the list structure shown in
Figure~\ref{fig:list1}. This means that when garbage collection is
run, the mark phase finds \lbminline{my-list} on the environment and
pushes this so-called root to the garbage collector stack. The mark
phase then runs for as long as there are values on its stack, in each
iteration popping and inspecting the top of the stack.  In this case,
the top of the stack is a pointer to the list structure \lbminline{(a
  b c)}.  The mark phase follows the pointer to the heap cell. If the
cell is already marked, this iteration is done and the next item on
the stack is popped. If the cell is not already marked, the mark bit
is set and the cdr field is pushed to the garbage collector stack. The
mark phase then goes back to the top with the car field as the new
value for the garbage collector to inspect.

In simplified C code that skips over many details of the LispBM mark
phase, the stack based process is implemented as follows:

\begin{ccode}
void lbm_gc_mark_phase(lbm_value root) {
  lbm_value t_ptr;
  lbm_stack_t *s = &lbm_heap_state.gc_stack;
  s->data[s->sp++] = root;

  while (!lbm_stack_is_empty(s)) {
    lbm_value curr;
    lbm_pop(s, &curr);

  mark_shortcut:

    if (!lbm_is_ptr(curr)) {
      continue;
    }

    lbm_cons_t *cell = &lbm_heap_state.heap[lbm_dec_ptr(curr)];

    if (lbm_get_gc_mark(cell->cdr)) {
      continue;
    }

    cell->cdr = lbm_set_gc_mark(cell->cdr);
    lbm_heap_state.gc_marked ++;

    lbm_push(s, cell->cdr);
    curr = cell->car;
    goto mark_shortcut; // Skip a push/pop
  }
}
\end{ccode}

One observation about the code above is that it favors cdr-leaning
values (see Figure~\ref{fig:list1}).  On lines 25 and 26, the cdr of the
current cell is pushed on the stack and the car value is immediately
handled. The garbage collector will always fully mark car subtrees
before popping and handling the cdr subtree. This means that as long
as values are mostly list-like (cdr-leaning), the garbage collector
will have no problem marking them, but it is also very easy to create
an adversarial car-leaning tree that exhausts the garbage collector
stack.

\begin{figure}
  \centering
  \includegraphics[width=0.9\linewidth]{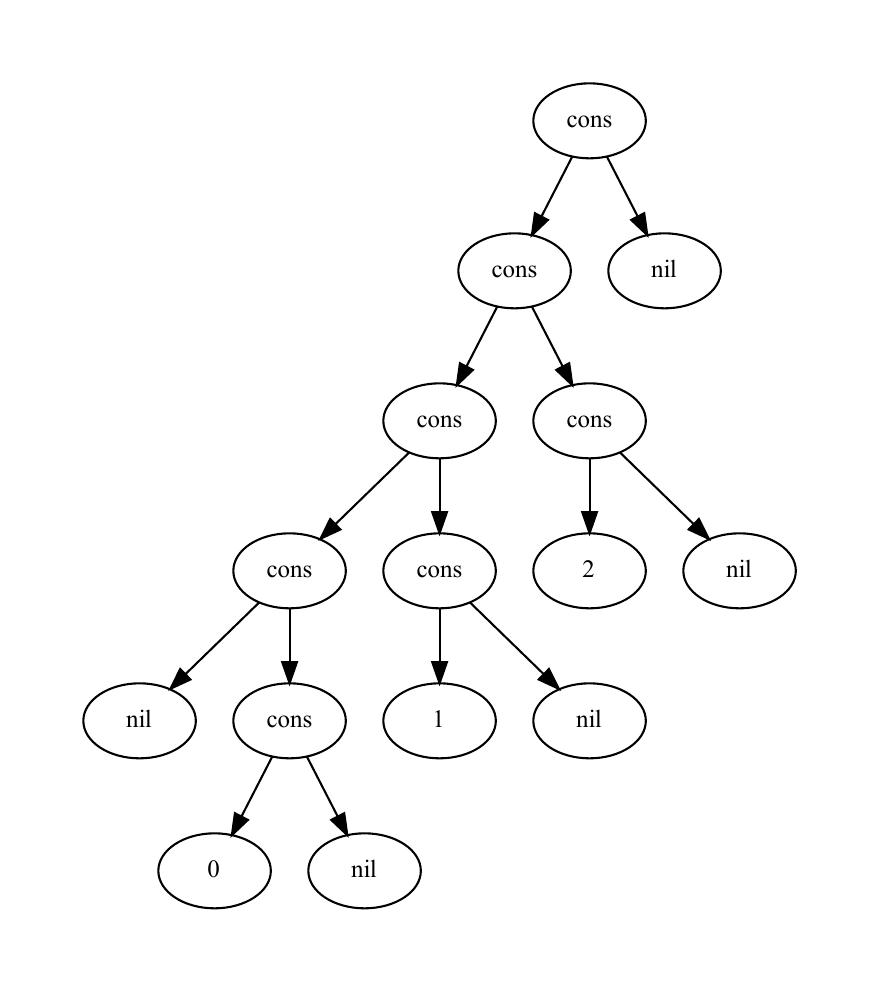}
  \caption{The heap structure built by \lbminline{(make-adversarial nil (range 3))}.}
  \label{fig:adversarial}
\end{figure}

The function below takes a list of elements and creates a car-leaning
adversarial value with the same values as the input list. The value
created by \lbminline{make-adversarial} is a snoc list with a pointer
to a heap cell in every car field along the leftwards spine; marking
this structure requires as deep a stack as the leftwards spine is
deep.  Figure~\ref{fig:adversarial} illustrates the shape of the
adversarial value.

\begin{lbmcode}
(defun make-adversarial (acc ls)
  (match ls
         (nil  (cons acc nil))
         (((? a) . (? b)) (make-adversarial (cons acc (cons a nil)) b))))
\end{lbmcode}

An adversary of 400 elements more than exhausts the garbage collector
stack. At the time, this was an unrecoverable error requiring a restart.

\begin{terminal}
# (define a (make-adversarial nil (range 400)))
> ((((((((((((((((((((((((((((((((((((...
# (gc)
# CRITICAL ERROR
ERROR: Critical Error
\end{terminal}

The solution to this problem comes in the form of the
Deutsch-Schorr-Waite pointer-reversal marking
algorithm~\cite{schorr1967efficient,knuth1997taocp1,jones1996garbage}.
The pointer-reversal marking algorithm uses only a constant amount of additional memory;
instead it rearranges pointers in the structure it is traversing.  The
trade-off here is that using the pointer-reversal algorithm each node
in a tree is visited three times, once on the way down a car-spine where
pointers are reversed, then again during cdr traversal, and lastly once more
to restore the pointers.

Up until recently, LispBM used the stack-based garbage collector by
default, and one had to compile with the \cinline{LBM\_USE\_GC\_PTR\_REV}
flag to get the pointer-reversal implementation. The extra accesses it
performs have a cost, shown in Table~\ref{tab:gc-stack-vs-ptrrev}, that
we did not want to pay unconditionally in the VESC firmware.

Even if the pointer-reversal algorithm costs more, it is more correct
and plugs a serious weakness in the runtime system. Most of the
performance hit can be avoided by using the stack-based mark phase as
the entry point and switching to pointer-reversal only once the stack is
full. Both mark phases are more complicated than shown here, and
combining them was a hairy operation needing extensive testing. This was
implemented in LispBM as of September 2026: the
\cinline{LBM\_USE\_GC\_PTR\_REV} flag is gone, pointer-reversal marking
is always compiled in, and the collector falls back to it automatically
when the stack is exhausted.

There is another notable garbage collection technique that could
potentially be applied to LispBM. This is the incremental sweep
algorithm~\cite{hughes1982semi} that, instead of sweeping the entire
heap in one go, sweeps just enough at the point of each allocation to
find a free cell. This is a very appealing and easy way to decrease
the garbage collector pause times, but unfortunately it is too
intrusive a change to the runtime system at this moment, and it is
very unclear what the total cost in performance would be. As LispBM
stores the mark bits in the values, if a mark bit can be set between
runs of the garbage collector, we could not compare two values without
first masking out the mark bit, for example. Many operations would get
the added cost of masking out mark bits to accommodate the incremental
sweep. Alternatively, if we break out all the mark bits into a
separate array, the incremental sweep could be implemented with no ill
effect. The separate array of mark bits, however, would cost a whole
extra memory access per cell touched by the garbage collector. This is
a complex solution space to explore.

\begin{table}
  \centering
  \caption{Average garbage-collection time for the stack-based and
    pointer-reversal mark phases, at two heap-fill levels, run on a
    Dash35b (VESC Express ESP32C3 based dashboard display) target with
    a 5120 cell heap.}
  \label{tab:gc-stack-vs-ptrrev}
  \footnotesize
  \begin{tabular}{lrrr}
    \toprule
    Heap usage & Stack-based (ms) & Pointer-reversal (ms) & Slowdown \\
    \midrule
    37\% & 1.5 & 2.1 & 1.4$\times$ \\
    83\% & 2.1 & 3.2 & 1.5$\times$ \\
    \bottomrule
  \end{tabular}
\end{table}

\section{Compacting Allocator}

In Section~\ref{sec:memory} the heap and \cinline{lbm\_memory} are
introduced.  The heap is immune to fragmentation as there is only a
one-size allocatable unit, the cell.  The \cinline{lbm\_memory},
however, allows allocation of arbitrary size arrays and is susceptible
to fragmentation. The fragmentation of the \cinline{lbm\_memory} is
not an issue as long as one allocates relatively small and relatively
uniformly sized objects.  But as data buffers allocated in
\cinline{lbm\_memory} increase in size the sensitivity to
fragmentation increases.

Motivated by a graphical user interface that used a number of large
graphics buffer objects, we started work on a defragmentable memory region
in the summer of 2024. To use this type of memory region the programmer
allocates a portion of the \cinline{lbm\_memory} using \lbminline{dm-create}:

\begin{lbmcode}
(define dm (dm-create 2048))
\end{lbmcode}

The code above creates a defragmentable memory area of 2048 bytes in
which we can allocate buffers using \lbminline{dm-alloc}. Many
functions, such as \lbminline{img-buffer} for creating graphical image
buffers, take an optional argument to indicate that the buffer should
be allocated in that defragmentable memory area.

Just like for any allocation in LispBM, if an allocation in a
defragmentable area fails then garbage collection runs and the
allocation is retried. In the case of a defragmentable memory
allocation, if the second attempt at allocation also fails, the
compaction process runs and then the allocation is attempted a third
time.

The compaction process moves buffers towards the beginning of the
defragmentable region until a hole opens up that is large enough for
the allocation to succeed.  Its doing just enough shuffling to find a
spot.

If buffers move in memory that means that references to those buffers
must be updated. Fortunately for us, \cinline{lbm\_memory}
allocations has a property that defragmentable allocations share,
there is never more than one reference to the allocated buffer!

The ``single-reference'' property comes from how every
\cinline{lbm\_memory} allocation has an associated heap-cell
representation (see Section~\ref{sec:memory}). Before we had
defragmentable areas, this associated heap-cell representation was a
way to make garbage collection (which only traverses the heap) aware
of the \cinline{lbm\_memory} allocation. Now, this heap-cell has the
additional role of being the single unique reference to the allocated
memory making the repointing of references to a moved array easy. The
compactible memory implementation keeps a back-pointer to the
heap-cell that refers to it in its allocation meta-data. The LispBM
runtime system never copies a cell that is acting like a sentinel for
a \lbminline{lbm\_memory} or compactible area array, and the type system
prohibits program-level inspection of the contents of such a cell.

\section{Supporting programs larger than the RAM heap}\label{sec:large}

\begin{figure}
  \centering
  \includegraphics[width=0.6\linewidth]{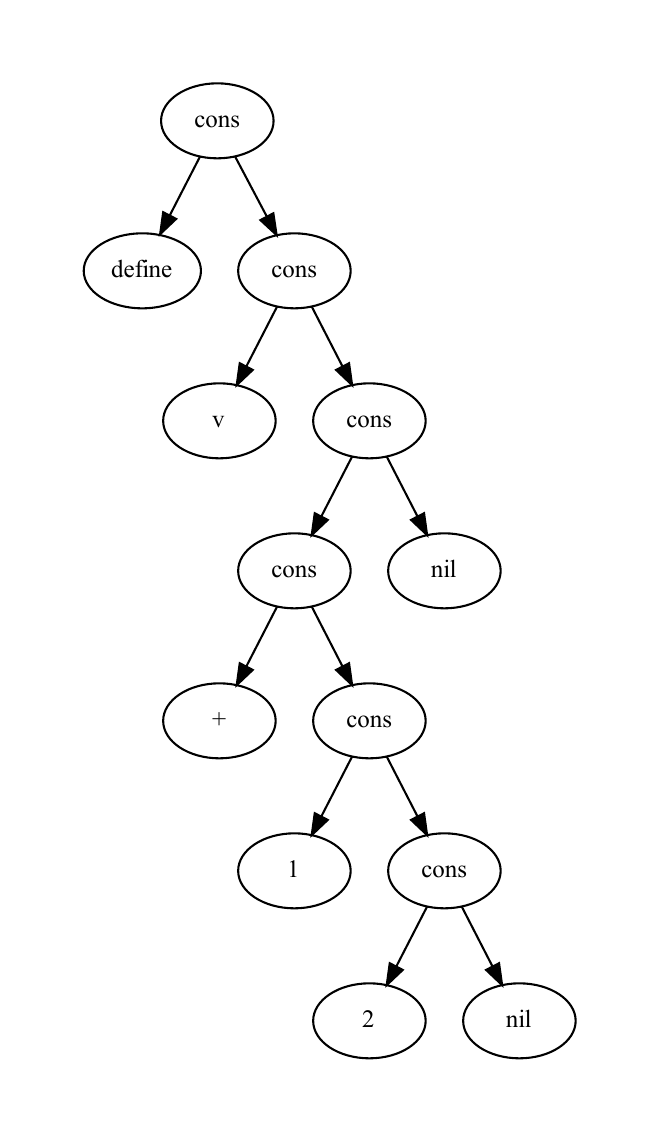}
  \caption{The heap structure created from reading the expression \lbminline{(define v (+ 1 2))}.}
  \label{fig:prog1}
\end{figure}

LispBM's starting point is to evaluate programs off the heap. That is,
the programs are read from source into a heap data structure that
represents the program.  For example, the program below is read into
the structure shown in Figure~\ref{fig:prog1}:

\begin{lbmcode}
(define v (+ 1 2))
\end{lbmcode}

From this, we can see that programs require heap space related to program size. 

\begin{figure}
  \centering
  \includegraphics[width=0.6\linewidth]{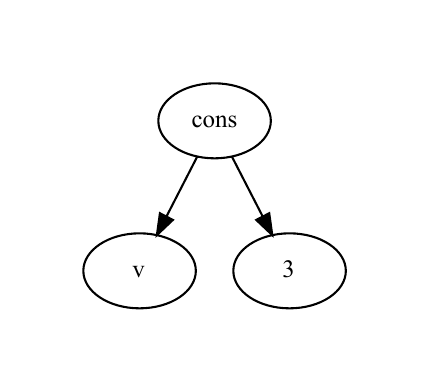}
  \caption{The heap structure holding binding of \lbminline{v} to the value of \lbminline{(+ 1 2)}.}
  \label{fig:prog-eval1}
\end{figure}

LispBM used to load the entire program source into a heap structure
before beginning to evaluate that structure. The first step taken to
reduce RAM usage was to implement incremental reading that evaluates
each top-level expression as soon as its heap-structure has been
formed. This approach reduces total RAM memory usage by replacing the
program above, \lbminline{(define v (+ 1 2))} with only its binding in
the global environment, see Figure~\ref{fig:prog-eval1}. In this
example it temporarily uses as many as 6 cells while reading and
evaluating, while only the one-cell environment binding contributes to
the total.

Adding the concept of incremental reading just bought us some time. The users soon caught up
and again wrote programs that exhausted all available heap or \cinline{lbm\_memory}.
To remedy this we added a way to move large bindings to flash, \lbminline{move-to-flash},
which allows the user to write:

\begin{lbmcode}
(define my-str "Hello world")
(move-to-flash my-str)
\end{lbmcode}

After the string has been moved to flash it is a constant string stored in flash memory:

\begin{terminal}
# (define my-str "Hello world")
> "Hello world"
# (constant? my-str)
> nil
# (move-to-flash my-str)
> t
# (constant? my-str)
> t
\end{terminal}

Code can also be moved to and evaluated from flash:

\begin{terminal}
# (define my-code '(print my-str))
> (print my-str)
# (move-to-flash my-code)
> t
# (eval my-code)
"Hello world"
> t
\end{terminal}

To accommodate the values that moves to flash, LispBM has what we call
a const-heap. It is an area of flash where we allocate cons-cells
as well as byte-arrays. This area in flash is never touched by the garbage
collector. The garbage collector knows to stay out of constant memory because of
a const-bit being set in references to this area of memory (this is the same bit that \lbminline{constant?}
checks).
The \lbminline{move-to-flash} operation traverses a value and duplicates its structure
on the const-heap. It then deletes the object from the RAM heap and repoints the binding
in the environment to the newly created const-heap value.

As a shorthand for the \lbminline{define} and \lbminline{move-to-flash} idiom we
introduce the constant blocks notation, \lbminline{@const-start} and \lbminline{@const-end}:

\begin{lbmcode}
@const-start
(define my-str "hello world")
(define my-code '(print my-str))
@const-end  
\end{lbmcode}

The code above results in the same constant string and code stored on flash as in the
previous examples.

Now, flash memory is a bit interesting in that it has some serious limitations.
First, one should try to not write to it too much because clearing and writing to
flash wears it out. Second, flash memory is usually only erasable in pretty large blocks,
the exact block size varies but it is more than byte granularity. Third, flash cannot be
arbitrarily updated at a byte granularity. What is possible is turning 1-bits to 0-bits
but the other way around requires an erase of a large block.

The peculiarities of flash memory meant that this \lbminline{move-to-flash} addition,
while helpful in making large programs, also added problems.

Note that at this point, LispBM still reloads the entire program from
source code on every boot-up. This means that the first time a program
that uses const-blocks runs, it creates a const-heap structure in
flash that cannot be altered locally at byte granularity. So, say that
you have a program that the first time it runs evaluates
\lbminline{(define my-str "Hello world")} and writes the string
``Hello world'' into flash memory at a specific address. Next time the
program boots and evaluates from source code, it must write the exact
same ``Hello world'' string to the exact same address in flash. That
is, programs must evaluate deterministically in relation to their flash
footprint. It is very easy to break this determinism property, for example:

\begin{lbmcode}
@const-start
(define my-val (list 1 2 3 (random)))
@const-end
\end{lbmcode}

The random value written into flash will make future reboots of this
program fail.  This is a limitation that is hard to explain to a user,
and it also appears in the most subtle of ways.

The \lbminline{move-to-flash} operation makes it possible to write
programs that are larger than the RAM heap. Instead, the limitation
becomes that no single top-level expression can be larger than the RAM
heap. It, however, also introduces a hard-to-explain determinism
requirement on programs, one that the programmer must uphold without any
direct help from the language or runtime system.

\section{Fast boot using Lisp images}

\urldef{\dashsrcurl}\url{https://github.com/vedderb/vesc_pkg/tree/2444a25e639c31c40f9c01cad1c530458979511c/dash35b}

At the end of the previous section, Section~\ref{sec:large}, the LispBM
runtime system is capable of running programs larger than its RAM heap. This
capability comes with an awkward limitation (determinism property) and
it still performs a full read of the entire source code and recreates the
complete state at every boot.

Some Lisp runtimes supply a \lbminline{(save-lisp-and-die)} function
that saves enough of the runtime state into an on disk binary
image. The image can then be used to restore the same state by
loading it into the runtime system. 

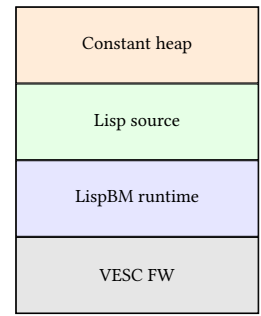
\begin{figure}
  \centering
  \begin{tikzpicture}[
      box/.style={draw, minimum width=3.2cm, minimum height=1cm, align=center, font=\footnotesize}
    ]
    \node[box, fill=orange!15]   (const)   {Constant heap};
    \node[box, fill=green!10, below=0pt of const]   (src)     {Lisp source};
    \node[box, fill=blue!10, below=0pt of src]      (rt)      {LispBM runtime};
    \node[box, fill=gray!20, below=0pt of rt]       (fw)      {VESC FW};

    \draw (const.north west) rectangle (fw.south east);
  \end{tikzpicture}
  \caption{Schematic layout of flash memory at boot. The VESC FW and
    LispBM runtime are fixed, pre-existing regions. The constant heap
    regions are written to flash the first time a program using
    \lbminline{move-to-flash} or constant blocks runs. Lisp source
    code can be written to flash using tooling associated with the
    VESC family of firmware.
  }
  \label{fig:flash-layout}
\end{figure}

We hypothesized that something like an image-system would improve the
boot-up time of LispBM systems. Actually going from idea to
implementation of this was quite complicated given the restrictions
of the target platforms.  We were already using RAM and big chunks of
flash memory for our programs and data, see
Figure~\ref{fig:flash-layout}. For example, we did not have enough
free flash to create full copies of the heap, \cinline{lbm\_memory}
and the constant heap; there is a symbol table and an extension table
as well. The idea to restore the full running state from a binary also poses
the question of what to do about peripherals that need initialization
or threads that are running in some particular state.

\begin{figure}
  \centering
  \begin{tikzpicture}[font=\footnotesize]
    % Outer Image region
    \draw (0,0) rectangle (3.2,4);
    \node[above] at (1.6,4.05) {Image};

    % Bootable image part, growing downward from the top
    \fill[blue!15] (0,2.8) rectangle (3.2,4);
    \draw (0,2.8) rectangle (3.2,4);
    \node at (1.6,3.4) {Bootable image};

    % Const heap, growing upward from the bottom
    \fill[orange!15] (0,0) rectangle (3.2,1.2);
    \draw (0,0) rectangle (3.2,1.2);
    \node at (1.6,0.6) {Const heap};

    % Growth direction arrows into the free space
    \draw[->, thick] (1.0,2.7) -- (1.0,2.1);
    \draw[->, thick] (2.2,1.3) -- (2.2,1.9);

    % Free space label
    \node[gray, font=\scriptsize\itshape] at (1.6,2.0) {free space};

    % Address markings
    \node[font=\scriptsize, anchor=west] at (3.3,4)   {High address};
    \node[font=\scriptsize, anchor=west] at (3.3,0)   {Low address};
  \end{tikzpicture}
  \caption{Schematic layout of an Image. The bootable image part
    grows downward from the top of the Image region, while the
    constant heap grows upward from the bottom. When the two meet
    somewhere in the middle, the image is full.}
  \label{fig:image-layout}
\end{figure}
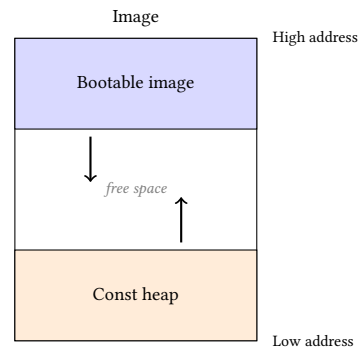

To solve the problems mentioned above, the Image system we designed
became a much more central part of LispBM's design. We require that
there is a region of memory provided by the integrator for a
Lisp-image. The so-called constant heap is created inside of this
memory area as shown in Figure~\ref{fig:image-layout}. This takes care
of the const-heap: it will always be a part of the image. But to restore
the state, the environment needs to be recreated. The environment can
contain pointers into the heap and the \cinline{lbm\_memory}. To avoid
backing up the entire heap and \cinline{lbm\_memory} we instead
traverse the environment and serialize all of its bindings into the image.

The image area is split into two regions as shown in
Figure~\ref{fig:image-layout}.  At the lowest image address, the
constant heap starts and it grows upwards.  At the top of the image
memory area, a region we call the ``bootable image'' starts.  The
bootable image is a sequence of entries that are read out one after
the other as the LispBM runtime system boots up. A binding from the
environment is added to the image as a {\em binding} entry containing
the symbol to bind to and the serialized (flattened) value it is bound
to. Symbols are another example: they are added to the image upon
creation as a symbol table entry (one link in the symbol table) as
well as a symbol string entry. While symbols are added to
the image at creation time, the environment is saved only when the
programmer calls a \lbminline{image-save} function. 

The problem of initializing peripherals and restoring running threads
was solved by entirely circumventing it and, plainly, deciding not to
even attempt it. Instead the programmer is expected to provide a
function called \lbminline{main} that initializes peripherals and
starts threads.  The main function is identified at
\lbminline{image-save} and a {\em start-up-function} entry is added to
the bootable region of the image. The typical form of a LispBM program
using images and a start-up, main, function is shown below:

\begin{lbmcode}
@const-start

(define a ...)
(define b ...)

(defun main () {

  (init-1)
  (init-2)

  (spawn thread1)
  (spawn thread2)

  })
@const-end

(image-save)
(main)
\end{lbmcode}

The program has a constant block with definitions of constants. It
defines a \lbminline{main} function which serves as the program entry
point. The \lbminline{main} function contains initializations of
hardware peripherals and spawns the main application threads.  Outside
of the constant block, the image is saved using
\lbminline{image-save}.

To perform the task of serializing the bindings of the environment we
enlisted the trusty \lbminline{flatten} function which at this time had been
with us for a long time.

\begin{terminal}
# (define a (list 1 2 3))
> (1 2 3)
# (define b (list a a a))
> ((1 2 3) (1 2 3) (1 2 3))
# (flatten b)
> [1 1 5 0 0 0 1 1 5 0 0 0 2 1 5 0 0 0 3 3 110 105 108 0 1 1 5 0 0 0 1 1 5 0 0 0 2 1 5 0 0 0 3 3 110 105 108 0 1 1 5 0 0 0 1 1 5 0 0 0 2 1 5 0 0 0 3 3 110 105 108 0 3 110 105 108 0]
\end{terminal}

The program above illustrates that as flatten traverses a list structure with sharing, such as \lbminline{(list a a a)},
each of the \lbminline{a} references is traversed and added independently to the flat value.
This is the easy way to implement a function like \lbminline{flatten} and while it is, of course, not entirely
correct, it is an acceptable trade-off when used mostly to send data over CAN to another unit that
does not know about the \lbminline{a} reference at all.

\begin{terminal}
# (unflatten (flatten b))
> ((1 2 3) (1 2 3) (1 2 3))
\end{terminal}

In the example above, the unflattened value is structurally different
from the \lbminline{b} used as input.

Now, when \lbminline{flatten} is used to store and restore the program environment, this
unfolding leads to an increase in size as well as a weird inconsistency in that values
change structurally when being restored after a reboot. The solution to this problem
is a redesign of how flattening works: we need to detect and track sharing. This led to
the creation of generalized pointer-reversal traversals that are run over the data structure
to flatten, first to detect sharing and build a sharing table, then again to generate
a flat representation with references to the sharing table. The sharing table is added
to the image and the exact structure of the value can be recreated upon reboot. 

For the Dash35b~\footnote{\dashsrcurl}, a vehicle dashboard application, the first boot that
creates the image takes 4.3 seconds. Successive boot-ups take 2 milliseconds. The first
boot is quite similar to how LispBM would have booted up that
application before introducing the image system, and at that time we
saw long startup times on every boot. Note that during the first boot,
the incremental reader runs: it reads and evaluates interleaved, and
it executes \lbminline{move-to-flash} on computed results. Starting a
saved image just restores the environment and other data structures
from the bootable image region and calls the registered start-up
function.

Now, the image is obligatory. Indeed, there must be an image for any
LispBM program to run, even one that makes no use of
\lbminline{image-save}, constant blocks, or a \lbminline{main}
function. There is, however, no requirement that the image is in flash. So, a limited
all-RAM LispBM integration is still very possible.

The image system also improves the situation in regards to the
determinism property. When using an image, the program is only
evaluated from source one time, so the entire problem of successive
boots needing to evaluate the source identically is circumvented.

\section{Related Work}

Making a Lisp of one's own is a rite of passage for programmers. We
focus here on the handful of Lisp-like languages implemented
specifically for microcontrollers and embedded systems.

uLisp~\cite{JohnsonDavies2026uLisp} is a microcontroller Lisp
targeting Arduino platforms. uLisp is a subset of Common Lisp, so,
given the right environment, uLisp programs should run on Common Lisp as
well. uLisp seems to aim to be portable so that uLisp programs can
be directly reused on different platforms supporting the runtime. This
differs from LispBM which does not try to be Lisp-program portable
across platforms but rather is a family of languages, one per
integration. In reality, there is a high degree of compatibility and a
big shared library for the BLDC (motor control) firmware and the
VESC Express firmware, but they are not entirely the same. A motor
controller and dashboard display have different needs when it
comes to scriptability. Differences in what functionality they
expose to the script therefore make sense.

Ferret~\cite{akkaya_ferret} is a Lisp dialect for real-time applications
on microcontrollers. When it comes to syntax and semantics, Ferret takes
inspiration from Clojure~\cite{Hickey2020Clojure}. Ferret compiles down
to C++ code that is then compiled for the target platform and deployed.
Here Ferret and LispBM are working towards opposite goals. LispBM is not
for guaranteed real-time software while Ferret is; LispBM enables systems
to be modified without recompiling the firmware binary and Ferret does not.
These choices are both valid.

Microscheme~\cite{suchocki_microscheme} implements a subset of Scheme
and targets Arduino platforms. Microscheme programs are compiled down
to assembler code that is then assembled into binary form for the
target. This makes Microscheme more similar to Ferret than to uLisp
and LispBM, which are mainly interpreted languages. Otherwise, in its
aims, Microscheme feels similar to uLisp in being a faithful implementation
of a standardized language.

If we go outside of Lisp-like languages, there is a very interesting
new Haskell compiler and runtime system called
MicroHs~\cite{Augustsson2026MicroHs}.  Krook and
Augustsson~\cite{krook2026cloudmicrohaskell} have been able to run
MicroHs-compiled code on microcontrollers with promising results,
though at the cost of high memory usage. MicroHs programs compile down
to a combinator graph that is evaluated on the microcontroller. While
evaluating, the graph is modified in place, making execution from
flash a challenge. Lisp-like languages allow modification of code, but
that is as part of program generation capability not as a required
part of evaluation. LispBM gives up the ability to modify code if that
code is stored in flash.

In the space of scripting languages for microcontrollers, there is one
behemoth, MicroPython~\cite{MicroPythonProject2026}. One of MicroPython's
strengths is that it shares the syntax and semantics of the highly popular
Python programming language. With this comes a naturally large and active
community of contributors and users, a situation that is easy to envy.
MicroPython shares, broadly, the application domain of the rest of the related
work here and, like LispBM, provides a read-evaluate-print-loop (REPL) on the
microcontroller.

\section{Discussion}

We are having a lot of fun implementing LispBM and its
libraries. The integration of LispBM into the VESC firmware brought
additional challenges and puzzles to solve. If author 1 had been left
alone to play with the LispBM implementation, then no one would ever
have reached the point where programs exhausted all RAM, for example,
and the peculiar but useful constant heap concept would never have
been designed. The constant heap led to the image system, and the
image system cast light on needed sharing-detecting, pointer-reversing
traversal functions. Those traversals were such a puzzle, and that's
what we like!

As LispBM was first integrated into the VESC firmware as an
experiment, we had cooperative concurrency and message passing, but
still no way to negate a number in a Lisp program!  This illustrates
clearly that up until that point, the main implementer of LispBM didn't
really care that much about writing programs in LispBM. It was writing
on LispBM itself that mattered. This was going to change
dramatically. Things moved quite fast in the beginning and we expanded
on functionality and documentation. A few users turned up and wrote
simple scripts to interface their motor controllers with existing
dashboards or similar.  There was some grumbling here and there about
Lisp and its weird syntax, but mostly people seemed genuinely happy
that there was now a scripting language in the motor controller.
Hobby makers and engineers seem to just pick up
whatever tools they are given and build. Taste in language is highly
subjective, but it is encouraging and fun to see one's tools picked up
and used in ways one could not imagine.

In this paper we have seen stories on how LispBM, after finding a home
in the VESC firmware, got a new concurrency model to simplify adoption
for a non-programmer userbase, how we experimented with garbage
collection to fix the collector's adversarial-input weakness, how the need for a compacting allocator arose
and how we solved it within our existing \cinline{lbm\_memory}
framework, how we manage to run programs that are much larger than our
RAM heap, and finally, how we speed up the boot process using a flavor
of Lisp images.

The garbage collector used to run on a stack with a very
easy-to-describe adversarial input; this has since been fixed by falling
back to pointer-reversal marking when the stack is exhausted (Section~\ref{sec:gc}).

Most recently, a lot of the work on LispBM has been focusing on its
libraries. We aim to have small, efficient libraries specifically made
for microcontrollers and the kind of applications users write. We are
happy to see a surge of helpful contributors in this area.

%\begin{acks}
%\end{acks}

%% ---------------------------------------------------------------------
\bibliographystyle{ACM-Reference-Format}
\bibliography{references}

@inproceedings{Krook2022Scoria,
  author    = {Krook, Robert and Hui, John and Svensson, Bo Joel and Edwards, Stephen A. and Claessen, Koen},
  title     = {Creating a Language for Writing Real-Time Applications for the {I}nternet of {T}hings},
  booktitle = {2022 20th ACM-IEEE International Conference on Formal Methods and Models for System Design (MEMOCODE)},
  year      = {2022},
  address   = {Shanghai, China},
  publisher = {ACM-IEEE},
  doi       = {10.1109/MEMOCODE57689.2022.9954383},
  isbn      = {9798350331905}
}

@inproceedings{Sarkar2021SenseVM,
  author    = {Sarkar, Abhiroop and Krook, Robert and Svensson, Bo Joel and Sheeran, Mary},
  title     = {Higher-Order Concurrency for Microcontrollers},
  booktitle = {Proceedings of the 18th ACM SIGPLAN International Conference on Managed Programming Languages and Runtimes (MPLR '21)},
  year      = {2021},
  address   = {M{\"u}nster, Germany},
  publisher = {ACM},
  doi       = {10.1145/3475738.3480716}
}

@inproceedings{Sarkar2022Synchron,
  author    = {Sarkar, Abhiroop and Svensson, Bo Joel and Sheeran, Mary},
  title     = {Synchron -- An {API} and Runtime for Embedded Systems},
  booktitle = {36th European Conference on Object-Oriented Programming (ECOOP 2022)},
  year      = {2022},
  series    = {LIPIcs},
  volume    = {222},
  pages     = {17:1--17:29},
  address   = {Dagstuhl, Germany},
  publisher = {Schloss Dagstuhl -- Leibniz-Zentrum f{\"u}r Informatik},
  doi       = {10.4230/LIPIcs.ECOOP.2022.17}
}

@inproceedings{Armstrong2007History,
  author    = {Armstrong, Joe},
  title     = {A History of {E}rlang},
  booktitle = {Proceedings of the Third ACM SIGPLAN Conference on History of Programming Languages (HOPL III)},
  year      = {2007},
  address   = {San Diego, CA, USA},
  publisher = {ACM},
  doi       = {10.1145/1238844.1238850}
}

@misc{Bazon2012CPSEvaluator,
  author       = {Bazon, Mihai},
  title        = {{CPS} Evaluator},
  howpublished = {\url{https://lisperator.net/pltut/cps-evaluator/}},
  year         = {2012},
  note         = {Part of the ``How to Implement a Programming Language'' tutorial series on Lisperator.net. Accessed 2026-08-10}
}

@misc{Svensson2022Defunctionalization,
  author       = {Svensson, Bo Joel},
  title        = {Defunctionalization of a Continuation-Passing Style Evaluator},
  howpublished = {\url{https://www.lispbm.com/pages/lispbm-evaluator/index.html}},
  year         = {2022},
  note         = {LispBM.com. Accessed 2026-08-10}
}

@misc{JohnsonDavies2026uLisp,
  author       = {Johnson-Davies, David},
  title        = {{uLisp}: Lisp for microcontrollers},
  howpublished = {\url{http://www.ulisp.com}},
  note         = {Official uLisp website. Accessed 2026-08-14}
}

@misc{MicroPythonProject2026,
  author       = {{MicroPython Project}},
  title        = {{MicroPython}: {Python} for microcontrollers},
  howpublished = {\url{https://micropython.org}},
  year         = {2026},
  note         = {Official MicroPython website. Accessed 2026-08-23}
}

@misc{Augustsson2026MicroHs,
  author       = {Augustsson, Lennart},
  title        = {{MicroHs}: {Haskell} implemented with combinators},
  howpublished = {\url{https://github.com/augustss/MicroHs}},
  note         = {Official MicroHs repository. Accessed 2026-08-23}
}

@article{schorr1967efficient,
  title={An efficient machine-independent procedure for garbage collection in various list structures},
  author={Schorr, Herbert and Waite, William M},
  journal={Communications of the ACM},
  volume={10},
  number={8},
  pages={501--506},
  year={1967},
  publisher={ACM New York, NY, USA}
}

@book{jones1996garbage,
  title={Garbage collection: algorithms for automatic dynamic memory management},
  author={Jones, Richard and Lins, Rafael},
  year={1996},
  publisher={John Wiley \& Sons, Inc.}
}

@article{hughes1982semi,
  title={A semi-incremental garbage collection algorithm},
  author={Hughes, R John M},
  journal={Software: Practice and Experience},
  volume={12},
  number={11},
  pages={1081--1082},
  year={1982},
  publisher={Wiley Online Library}
}

@book{Steele1990CommonLisp,
  author    = {Guy Lewis Steele Jr.},
  title     = {Common LISP: The Language},
  edition   = {2nd},
  year      = {1990},
  publisher = {Digital Press},
  address   = {Woburn, MA, USA},
  isbn      = {1-55558-041-6}
}

@techreport{SussmanSteele1975,
  author      = {Gerald Jay Sussman and Guy Lewis Steele Jr.},
  title       = {{SCHEME}: An Interpreter for Extended Lambda Calculus},
  institution = {Massachusetts Institute of Technology},
  type        = {AI Memo},
  number      = {349},
  address     = {Cambridge, Massachusetts},
  month       = {December},
  year        = {1975}
}

@article{mccarthy1960recursive,
  title={Recursive functions of symbolic expressions and their computation by machine, part I},
  author={McCarthy, John},
  journal={Communications of the ACM},
  volume={3},
  number={4},
  pages={184--195},
  year={1960},
  publisher={ACM New York, NY, USA},
  doi={10.1145/367177.367199}
}

@book{knuth1997taocp1,
  author = {Knuth, Donald E.},
  title = {The Art of Computer Programming, Volume 1: Fundamental Algorithms},
  edition = {3rd},
  year = {1997},
  publisher = {Addison-Wesley},
  address = {Reading, Massachusetts},
  isbn = {978-0201896831}
}

@misc{akkaya_ferret,
  author       = {Nurullah Akkaya},
  title        = {{Ferret Programmer's Manual: A hard real-time Clojure for Lisp machines}},
  howpublished = {\url{https://ferret-lang.org}},
  note         = {Accessed: 2026-08-23}
}

@misc{suchocki_microscheme,
  author       = {Ryan Suchocki},
  title        = {{Microscheme: A Scheme subset for Atmel 8-bit AVR microcontrollers}},
  howpublished = {\url{https://github.com/ryansuchocki/microscheme}},
  note         = {Accessed: 2026-08-23}
}

@inproceedings{krook2026cloudmicrohaskell,
  author = {Krook, Robert and Augustsson, Lennart},
  title = {{CloudMicroHaskell}: Direct-Style Distributed {H}askell via Runtime Graph Serialisation},
  booktitle = {Haskell '26: Proceedings of the 19th ACM SIGPLAN International Haskell Symposium},
  pages = {34--64},
  year = {2026},
  month = aug,
  publisher = {Association for Computing Machinery},
  address = {New York, NY, USA},
  doi = {10.1145/3830439.3831272}
}

@article{Hickey2020Clojure,
  author = {Hickey, Rich},
  title = {A History of Clojure},
  year = {2020},
  issue_date = {June 2020},
  publisher = {Association for Computing Machinery},
  address = {New York, NY, USA},
  volume = {4},
  number = {HOPL},
  doi = {10.1145/3386321},
  journal = {Proc. ACM Program. Lang.},
  month = {jun},
  articleno = {71}
}

\end{document}